\documentstyle[12pt]{article}
\newcommand{\n}{\hspace*{-2.5mm}}
\newcommand{\gsim}{\;\rlap{\lower 3.5 pt \hbox{$\mathchar \sim$}} \raise 1pt
 \hbox {$>$}\;}
\newcommand{\lsim}{\;\rlap{\lower 3.5 pt \hbox{$\mathchar \sim$}} \raise 1pt
 \hbox {$<$}\;}
\newcommand{\re}{\mathop{\rm Re}\nolimits}
\newcommand{\im}{\mathop{\rm Im}\nolimits}
\newcommand{\di}{\mathop{{\rm Li}_2}\nolimits}
\renewcommand{\O}{{\cal O}}
\catcode`@=11
\newcount\@tempcntc
\def\@citex[#1]#2{\if@filesw\immediate\write\@auxout{\string\citation{#2}}\fi
  \@tempcnta\z@\@tempcntb\m@ne\def\@citea{}\@cite{\@for\@citeb:=#2\do
    {\@ifundefined
       {b@\@citeb}{\@citeo\@tempcntb\m@ne\@citea\def\@citea{,}{\bf ?}\@warning
       {Citation `\@citeb' on page \thepage \space undefined}}%
    {\setbox\z@\hbox{\global\@tempcntc0\csname b@\@citeb\endcsname\relax}%
     \ifnum\@tempcntc=\z@ \@citeo\@tempcntb\m@ne
       \@citea\def\@citea{,}\hbox{\csname b@\@citeb\endcsname}%
     \else
      \advance\@tempcntb\@ne
      \ifnum\@tempcntb=\@tempcntc
      \else\advance\@tempcntb\m@ne\@citeo
      \@tempcnta\@tempcntc\@tempcntb\@tempcntc\fi\fi}}\@citeo}{#1}}
\def\@citeo{\ifnum\@tempcnta>\@tempcntb\else\@citea\def\@citea{,}%
  \ifnum\@tempcnta=\@tempcntb\the\@tempcnta\else
   {\advance\@tempcnta\@ne\ifnum\@tempcnta=\@tempcntb \else \def\@citea{--}\fi
    \advance\@tempcnta\m@ne\the\@tempcnta\@citea\the\@tempcntb}\fi\fi}
\catcode`@=12
\begin{document}
\title{\vskip-3cm{\baselineskip14pt
\leftline{\normalsize\hspace*{13cm}\bf KEK--TH--412\hfill}
\leftline{\normalsize\hspace*{13cm}\bf September 1994\hfill}
}
\vskip1.5cm
Status of Higher-Order Corrections in the Standard Electroweak
Theory\thanks{To appear in {\it Int.\ J. Mod.\ Phys.\ A\/}.}
}
\author{Bernd A. Kniehl\thanks{On leave from
{\it II. Institut f\"ur Theoretische Physik, Universit\"at Hamburg,
Luruper Chaussee 149, 22761 Hamburg, Germany\/}; address after 1 October 1994:
{\it Max-Planck-Institut f\"ur Physik, F\"ohringer Ring 6, 80805 Munich,
Germany\/}.}\ \thanks{JSPS Fellow.}\\
Theory Division, National Laboratory for High Energy
Physics (KEK),\\
1-1 Oho, Tsukuba-shi, Ibaraki-ken, 305 Japan}
\date{}
\maketitle
\begin{abstract}
We review recent theoretical progress in the computation of
radiative corrections beyond one loop within the standard model of
electroweak interactions, both in the gauge and Higgs sectors.
In the gauge sector, we discuss universal corrections of
$\O(G_F^2M_H^2M_W^2)$, $\O(G_F^2m_t^4)$,
$\O(\alpha_sG_FM_W^2)$, $\O(\alpha_s^2G_Fm_t^2)$, and those due to virtual
$t\bar t$ threshold effects, as well as specific corrections to
$\Gamma\left(Z\to b\bar b\right)$ of
$\O(G_F^2m_t^4)$, $\O(\alpha_sG_F$\break $m_t^2)$, and
$\O(\alpha_s^3)$ including finite-$m_b$ effects.
We also present an update of the hadronic contributions to $\Delta\alpha$.
Theoretical uncertainties, other than those due to the lack of knowledge of
$M_H$ and $m_t$, are estimated.
In the Higgs sector, we report on the $\O(\alpha_sG_Fm_t^2)$ corrections to
$\Gamma\left(H\to f\bar f\,\right)$ including those which are specific for the
$b\bar b$ mode, the $\O(\alpha_s^2)$ corrections to
$\Gamma\left(H\to q\bar q\right)$ including the finite-$m_q$ terms, and the
$\O(\alpha_s)$ corrections to $\Gamma(H\to gg)$.
\end{abstract}

\section{Introduction}

Precision tests of the standard electroweak theory \cite{gsw} have reached
a mature stage since their beginnings more than two decades ago.
This topic is routinely on the agenda of the major high-energy-physics
conferences and workshops \cite{con}, and numerous comprehensive---and in
some cases even encyclopaedic---reviews on the status of quantum corrections
have been published in the recent few years \cite{pub,hol,bur,jeg,hal,bak}.
In this article, an attempt is made to summarize the latest developments in
the study of higher-order corrections to the electroweak parameters and the
partial decay widths of the $Z$ and Higgs bosons in the framework of the
minimal standard model.
In the case of the Higgs boson, we shall concentrate on the low- and
intermediate mass range, $M_H<2M_W$, which will be accessed by
colliding-beam experiments in the near future.

As a rule, the size of radiative corrections to a given process
is determined by the discrepancy between the various mass and
energy scales involved.
In $Z$-boson physics, the dominant effects arise from light
charged fermions, which induce large logarithms of the form
$\alpha^n\ln^m(M_Z^2/m_f^2)$ $(m\le n)$ in the
fine-structure constant (and also in initial-state radiative
corrections), and from the top quark, which generates power corrections
of the orders $G_Fm_t^2$, $G_F^2m_t^4$, $\alpha_sG_Fm_t^2$, etc.
On the other hand, the quantum effects due to a heavy Higgs boson
are screened, i.e., logarithmic in $M_H$ at one loop and just quadratic
at two loops.
By contrast, such corrections are proportional to $M_H^2$ and $M_H^4$,
respectively, in the Higgs sector.

\section{Gauge Sector}

\subsection{Universal Corrections: Electroweak Parameters (Oblique\break
Corrections)}

For a wide class of low-energy and $Z$-boson observables,
the dominant effects originate entirely in the
gauge-boson propagators (oblique corrections)
and may be parametrized conveniently in terms of
four electroweak parameters, $\Delta\alpha$, $\Delta\rho$, $\Delta r$,
and $\Delta\kappa$, which bear the following physical meanings \cite{bur}:
\begin{enumerate}
\item $\Delta\alpha$ determines the running fine-structure constant at the
$Z$-boson scale, $\alpha(M_Z)/\alpha=(1-\Delta\alpha)^{-1}$,
where $\alpha$ is the corresponding value at the electron scale;
\item $\Delta\rho$ measures the quantum corrections to the ratio
of the neutral- and charged-current amplitudes at low energy,
$G_{NC}(0)/G_{CC}(0)=(1-\Delta\rho)^{-1}$ \cite{ros};
\item $\Delta r$ embodies the non-photonic corrections to the muon
lifetime,
$G_F=\left(\pi\alpha/\sqrt2s_w^2\right.$\break
$\left.\vphantom{\sqrt2}M_W^2\right)(1-\Delta r)^{-1}$ \cite{sir};
\item $\Delta\kappa$ controls the effective weak mixing angle,
$\bar s_w^2=s_w^2(1+\Delta\kappa)$,
that occurs in the ratio of the $f\bar fZ$ vector and axial-vector
couplings, $v_f/a_f=1-4|Q_f|\bar s_w^2$ \cite{hol}.
\end{enumerate}
Unless stated otherwise, we adopt the on-shell scheme \cite{aok} and set
$c_w^2=1-s_w^2=M_W^2/M_Z^2$ \cite{sir}.
The large logarithms are collected by $\Delta\alpha$, and the leading $m_t$
dependence is carried by $\Delta\rho$.
$\Delta r$ and $\Delta\kappa$ may be decomposed as
$(1-\Delta r)=(1-\Delta\alpha)(1+c_w^2/s_w^2\Delta\rho)-\Delta r_{rem}$
and $\Delta\kappa=c_w^2/s_w^2\Delta\rho+\Delta\kappa_{rem}$, respectively,
where the remainder parts are devoid of $m_f$ logarithms and $m_t$
power terms.
The triplet $(\Delta\rho,\Delta r_w,\Delta\kappa)$, where
$\Delta r_w$ is defined by
$(1-\Delta r)=(1-\Delta\alpha)(1-\Delta r_w)$,
is equivalent to synthetical sets like $(S,T,U)$ and
$(\varepsilon_1,\varepsilon_2,\varepsilon_3)$ \cite{pes},
which have gained vogue recently.
We note in passing that the bosonic contributions to these electroweak
parameters are, in general, gauge dependent and finite only in a restricted
class of gauges if the conventional formulation in terms of vacuum
polarizations is employed.
This problem may be cured in the framework of the pinch technique \cite{deg}.

At two loops, large contributions are expected to arise from
the exchange of heavy Higgs bosons, heavy top quarks, and gluons.
The hadronic contributions to $\Delta\alpha$ and the
$t\bar t$ threshold effects on $\Delta\rho$, $\Delta r$,
and $\Delta\kappa$ cannot be calculated reliably in
QCD to finite order.
However, they may be related via dispersion relations to
data of $e^+e^-\to hadrons$ and theoretical predictions of
$e^+e^-\to t\bar t$ based on realistic quark potentials, respectively.

\vglue 0.3cm
\leftline{\it 2.1.1 Two-Loop $\O(G_F^2M_H^2M_Z^2)$ Corrections}
\vglue 1pt

Such corrections are generated by two-loop gauge-boson vacuum-polarization
diagrams that are constructed from physical and unphysical Higgs bosons.
Knowledge \cite{bij} of the first two terms of the Taylor expansion around
$q^2=0$ is sufficient to derive the leading contributions to
$\Delta\rho$, $\Delta r$, and $\Delta\kappa$ \cite{hal,bcs},
\begin{eqnarray}
\Delta\rho&\n=\n&{G_F^2M_H^2M_W^2\over64\pi^4}\,{s_w^2\over c_w^2}
\left[-9\sqrt3\di\left({\pi\over3}\right)+{9\over2}\zeta(2)
+{9\over4}\pi\sqrt3-{21\over8}\right]\nonumber\\
&\n\approx\n&4.92\cdot10^{-5}\left({M_H\over1\,{\rm TeV}}\right)^2,
\nonumber\\
\Delta r&\n=\n&{G_F^2M_H^2M_W^2\over64\pi^4}
\left[9\sqrt3\di\left({\pi\over3}\right)-{25\over18}\zeta(2)
-{11\over4}\pi\sqrt3+{49\over72}\right]\nonumber\\
&\n\approx\n&-1.05\cdot10^{-4}\left({M_H\over1\,{\rm TeV}}\right)^2,
\nonumber\\
\Delta\kappa&\n=\n&{G_F^2M_H^2M_W^2\over64\pi^4}
\left[-9\sqrt3\di\left({\pi\over3}\right)+{53\over18}\zeta(2)
+{5\over2}\pi\sqrt3-{119\over72}\right]\nonumber\\
&\n\approx\n&1.37\cdot10^{-4}\left({M_H\over1\,{\rm TeV}}\right)^2,
\end{eqnarray}
Due to the smallness of the prefactors, these contributions are insignificant
for $M_H\lsim1\,$TeV.

\vglue 0.3cm
\leftline{\it 2.1.2 Two-Loop $\O(G_F^2m_t^4)$ Corrections for
$M_H\ne0$}
\vglue 1pt

Also at two loops, $\Delta\rho$ picks up the leading large-$m_t$ term,
and $\Delta r$ and $\Delta\kappa$ depend on $m_t$ chiefly via $\Delta\rho$.
Neglecting $m_b$ and defining $x_t=\left(G_Fm_t^2/8\pi^2\sqrt2\right)$, one
has
\begin{equation}
\label{drho}
\Delta\rho=3x_t\left[1
+x_t\rho^{(2)}\left({M_H\over m_t}\right)
-{2\over3}\left(2\zeta(2)+1\right){\alpha_s(m_t)\over\pi}\right],
\end{equation}
where, for completeness, also the well-known $\O(\alpha_sG_Fm_t^2)$
term \cite{djo,kni} is included.
Very recently, also the $\O(\alpha_s^2G_Fm_t^2)$ term has been computed
\cite{avd}, the result being $(-21.27063+1.78621\,N_F)(\alpha_s/\pi)^2$,
where $N_F$ is the number of active quark flavours.
The contribution to this term due to the so-called double-triangle diagram was
originally calculated in Ref.~\cite{ans}.
The coefficient $\rho^{(2)}(r)$ \cite{bar,dfg} is negative for all plausible
values of $r$, bounded from below by $\rho^{(2)}(5.72)=-11.77$,
and exhibits the following asymptotic behaviour:
\begin{equation}
\rho^{(2)}(r)=
\cases{
\displaystyle
-12\zeta(2)+19-4\pi r+\O(r^2\ln r),
&if $r\ll1;$\cr
\displaystyle
6\ln^2r-27\ln r+6\zeta(2)+{49\over4}
+O\left({\ln^2r\over r^2}\right),
&if $r\gg1.$\cr}
\end{equation}
The value at $r=0$ \cite{hoo} greatly underestimates the effect.
Both $\O(G_F^2m_t^4)$ and $\O(\alpha_sG_Fm_t^2)$ corrections
screen the one-loop result and thus increase the value of $m_t$ predicted
indirectly from global analyses of low-energy, $M_W$, LEP/SLC, and other
high-precision data.
Recently, a first attempt was made to control subleading corrections to
$\Delta\rho$, of $\O\Bigl(G_F^2m_t^2M_Z^2\ln(M_Z^2/m_t^2)\Bigr)$, in an
SU(2) model of weak interactions, and significant effects were
found \cite{dfg}.

\vglue 0.3cm
\leftline{\it 2.1.3 Two-Loop $\O(\alpha_sG_FM_W^2)$ Corrections}
\vglue 1pt

For $m_t\gg M_W$, the bulk of the QCD corrections is concentrated in
$\Delta\rho$ [see Eq.~(\ref{drho})].
However, for realistic values of $m_t$, the subleading terms,
of $\O(\alpha_sG_FM_W^2)$, are significant numerically,
e.g., they amount to 20\% of the full two-loop QCD correction to
$\Delta r$ at $m_t=150$~GeV.
Specifically, one has \cite{hal,kni}
\begin{eqnarray}
\Delta r_{rem}&\n=\n&{G_FM_W^2\over\pi^3\sqrt2}\left\{
-\alpha_s(M_Z)\left({c_w^2\over s_w^2}-1\right)\ln c_w^2
+\alpha_s(m_t)\left[\left({1\over3}-{1\over4s_w^2}\right)
\ln{m_t^2\over M_Z^2}+A+{B\over s_w^2}\right]\right\},\nonumber\\
\Delta\kappa_{rem}&\n=\n&{G_FM_W^2\over\pi^3\sqrt2}\left\{
\alpha_s(M_Z){c_w^2\over s_w^2}\ln c_w^2
-\alpha_s(m_t)\left[\left({1\over6}-{1\over4s_w^2}\right)
\ln{m_t^2\over M_Z^2}+{A\over2}+{B\over s_w^2}\right]\right\},
\end{eqnarray}
where terms of $\O(M_Z^2/m_t^2)$ are omitted within the square brackets and
\begin{eqnarray}
A&\n=\n&{1\over3}\left[-4\zeta(3)+{4\over3}\zeta(2)+{5\over2}\right]
\approx-0.03833,\nonumber\\
B&\n=\n&\zeta(3)-{2\over9}\zeta(2)-{1\over4}
\approx0.58652.
\end{eqnarray}
For contributions due to the $tb$ doublet, $\mu=m_t$ is the
natural renormalization scale for $\alpha_s(\mu)$.

\vglue 0.3cm
\leftline{\it 2.1.4 Hadronic Contributions to $\Delta\alpha$}
\vglue 1pt

Jegerlehner has updated his 1990 analysis \cite{jeg} of the
hadronic contributions to $\Delta\alpha$
by taking into account the hadronic resonance parameters
specified in the 1992 report \cite{hik} by the Particle Data Group
and recently published low-energy $e^+e^-$ data taken at Novosibirsk.
The (preliminary) result at $\sqrt s=91.175$~GeV reads \cite{fje}
\begin{equation}
\label{dalp}
\Delta\alpha_{hadrons}=0.0283\pm0.0007,
\end{equation}
i.e., the central value has increased by $1\cdot10^{-4}$, while
the error has decreased by $\pm2\cdot10^{-4}$.
The latter is particularly important, since this error has long
constituted the dominant uncertainty for theoretical predictions of
electroweak parameters.
For comparison, we list the leptonic contribution up to two loops in
QED \cite{kni},
\begin{eqnarray}
\Delta\alpha_{leptons}&\n=\n&{\alpha\over3\pi}\sum_\ell
\left\{\ln{M_Z^2\over m_\ell^2}-{5\over3}
+{\alpha\over\pi}\left[{3\over4}\ln{M_Z^2\over m_\ell^2}
+3\zeta(2)-{5\over8}\right]+O\left({m_\ell^2\over M_Z^2}\right)\right\}
\nonumber\\
&\n=\n&0.031\,496\,6\pm0.000\,000\,4,
\end{eqnarray}
where the error stems from the current $m_\tau$ world average,
$m_\tau=(1777.0\pm0.4)$~MeV \cite{wei}.

\vglue 0.3cm
\leftline{\it 2.1.5 $t\bar t$ Threshold Effects}
\vglue 1pt

Although loop amplitudes involving the top quark are mathematically well
behaved, it is evident that interesting and possibly significant features
connected with the $t\bar t$ threshold cannot be accommodated when the
perturbation series is truncated at finite order.
In fact, perturbation theory up to $\O(\alpha\alpha_s)$ predicts a
discontinuous steplike threshold behaviour for
$\sigma\left(e^+e^-\to t\bar t\,\right)$.
A more realistic description includes the formation of toponium resonances
by multi-gluon exchange.
For $m_t\gsim130$~GeV, the revolution period of a $t\bar t$ bound state
exceeds its lifetime, and the individual resonances are smeared out to a
coherent structure.
By Cutkosky's rule \cite{cut}, $\sigma\left(e^+e^-\to t\bar t\,\right)$
corresponds to the absorptive parts of the photon and $Z$-boson
vacuum polarizations, and its enhancement at threshold induces additional
contributions in the corresponding real parts,
which can be computed via dispersive techniques.
Decomposing the vacuum-polarization tensor generated by the insertion
of a top-quark loop into a gauge-boson line as
\begin{equation}
\Pi_{\mu\nu}^{V,A}(q)=\Pi^{V,A}(q^2)g_{\mu\nu}+\lambda^{V,A}(q^2)q_\mu q_\nu,
\end{equation}
where $V$ and $A$ label the vector and axial-vector components and
$q$ is the external four-momentum, and imposing Ward identities,
one derives the following set of dispersion relations \cite{ks,kns}:
\begin{eqnarray}
\Pi^V(q^2)&\n=\n&{q^2\over\pi}\int{ds\over s}\,
{\im\Pi^V(s)\over q^2-s-i\epsilon},\nonumber\\
\Pi^A(q^2)&\n=\n&{1\over\pi}\int ds
\left[{\im\Pi^A(s)\over q^2-s-i\epsilon}+\im\lambda^A(s)\right].
\end{eqnarray}
The alternative set of dispersion relations proposed in Ref.~\cite{cgn}
does not, in general, yield correct results, as has been demonstrated
by establishing a perturbative counterexample, namely the
$\O(\alpha_sG_Fm_t^2)$ corrections to $\Gamma(H\to\ell^+\ell^-)$
(see Sect.~3.1.1) \cite{hll}.
In Ref.~\cite{tak}, it has been suggested that this argument may be extended
to all orders in $\alpha_s$ by means of the operator product expansion.
In the threshold region, only $\im\Pi^V(q^2)$ and $\im\lambda^A(q^2)$ receive
significant contributions and are related by
$\im\lambda^A(q^2)\approx-\im\Pi^V(q^2)/q^2$,
while $\im\Pi^A(q^2)$ is strongly suppressed due to centrifugal barrier
effects \cite{kns}.
Of course, $\lambda^V(q^2)=-\Pi^V(q^2)/q^2$ by the relevant Ward
identity \cite{ks}.
These contributions in turn lead to shifts in $\Delta\rho$, $\Delta r$,
and $\Delta\kappa$.
A crude estimation may be obtained by setting
$\im\Pi^V(q^2)=\im\Pi^V(4m_t^2)=\alpha_sm_t^2$
in the interval $(2m_t-\Delta)^2\le q^2\le4m_t^2$, where $\Delta$
may be regarded as the binding energy of the 1S state.
This yields
\begin{eqnarray}
\Delta\rho&\n=\n&-{G_F\over2\sqrt2}\,{\alpha_s\over\pi}m_t\Delta,
\nonumber\\
\Delta r&\n=\n&-{c_w^2\over s_w^2}\Delta\rho\left[1-
\left(1-{8\over3}s_w^2\right)^2{M_Z^2\over4m_t^2-M_Z^2}
+{16\over9}s_w^4{M_Z^2\over m_t^2}\right],\nonumber\\
\Delta\kappa&\n=\n&{c_w^2\over s_w^2}\Delta\rho\left[1-
\left(1-{8\over3}s_w^2\right){M_Z^2\over4m_t^2-M_Z^2}\right].
\end{eqnarray}
Obviously, the threshold effects have the same sign as the
$\O(\alpha_sG_Fm_t^2)$ corrections.
For realistic quark potentials, one has approximately $\Delta\propto m_t$,
so that the threshold contributions scale like $m_t^2$.
Again, $\Delta\rho$ is most strongly affected, while the corrections to
$\Delta r_{rem}$ and $\Delta\kappa_{rem}$ are suppressed by $M_Z^2/m_t^2$.
A comprehensive numerical analysis may be found in Refs.~\cite{kns,fan,nut}.
For 150~GeV${}\le m_t\le{}$200~GeV, the threshold effects enhance the QCD
corrections by roughly 30\%.

We emphasize that the above QCD corrections come with both experimental and
theoretical errors.
In the following, we shall estimate them for $\Delta\rho$
assuming $m_t=174$~GeV \cite{abe};
the results are similar for other universal electroweak parameters.
The experimental errors are governed by the $\alpha_s$ measurement,
$\alpha_s(M_Z)=0.118\pm0.006$ \cite{bet}.
This amounts to errors of $\pm5\%$ and $\pm18\%$ on the continuum and
threshold contributions to $\Delta\rho$, respectively.
This reflects the fact the $\alpha_s$ dependence is linear in the continuum,
while that of 1S peak height is approximately cubic.
Since these errors are strongly correlated, we add them linearly to obtain
an overall experimental error of $\pm8\%$ on the QCD correction to
$\Delta\rho$.
Theoretical errors are due to unknown higher-order corrections.
In the continuum, they are usually estimated by varying the renormalization
scale, $\mu$, of $\alpha_s(\mu)$ in the range $m_t/2\le\mu\le2m_t$,
which amounts to $\pm11\%$.
The theoretical error on the threshold contribution is mainly due to
model dependence and is estimated to be $\pm20\%$ by comparing conventional
quark potentials.
For $m_t$ as large as 174~GeV, there is also a minor uncertainty connected with
the choice of the lower bound of the dispersion integral, which, in
Ref.~\cite{kns}, was chosen to be $s=(2m_t-\Delta)^2$ with
$\Delta=\sqrt{m_t\Gamma_t}$, which is 15.6~GeV for $m_t=174$~GeV.
Variation of $\Delta$ in the wide range between $\sqrt{m_t\Gamma_t}/2$ and
$2\sqrt{m_t\Gamma_t}$ induces a shift of $\pm9\%$ in the threshold
contribution.
Combining the common experimental error and the theoretical errors
quadratically, we obtain a total error of $\pm12\%$ on the QCD correction to
$\Delta\rho$.
This corresponds to an absolute QCD-related uncertainty in $\Delta\rho$ of
$\pm1.5\cdot10^{-4}$.
Due to the magnification factor $c_w^2/s_w^2$, the corresponding error
on $\Delta r$ and $\Delta\kappa$ is $\pm5.0\cdot10^{-4}$.
We stress that, in the case of $\Delta r$ and thus the $M_W$ prediction
from the muon lifetime, this error is almost as large as the one from
hadronic sources introduced via $\Delta\alpha$ [see Eq.~(\ref{dalp})].
For higher $m_t$ values, it may even be larger.

In Eq.~(\ref{drho}), we have evaluated the $\O(\alpha_sG_Fm_t^2)$
correction at $\mu=m_t$, since this is the only scale available.
However, this is a leading-order QCD prediction, which suffers from the
usual scale ambiguity.
We may choose $\mu=\xi m_t$ in such a way that the
$\O\Bigl(\alpha_s(\mu)G_Fm_t^2\Bigr)$ calculation agrees with the
$\O\Bigl(\alpha_s(m_t)G_Fm_t^2\Bigr)$ one plus the $t\bar t$ threshold
effects.
In the case of $\Delta\rho$, this leads to $\xi=0.190{+0.097\atop-0.057}$,
where we have included a $\pm30\%$ error on the $t\bar t$ threshold
contribution.
Alternative, conceptually very different approaches of scale setting
\cite{sv,asi,alb} yield results in the same ball park.
In Ref.~\cite{sv}, it is suggested that long-distance effects lower
the renormalization point for $\alpha_s(\mu)$ in Eq.~(\ref{drho}) through the
contributions of the near-mass-shell region to the evolution of the quark mass
from the mass shell to distances of order $1/m_t$.
To estimate these effects, the authors of Ref.~\cite{sv} apply the
Brodsky-Lepage-Mackenzie (BLM) criterion \cite{blm} to Eq.~(\ref{drho}) and
find $\xi=0.154$.
The author of Ref.~\cite{asi} expresses first the fermionic contribution to
$\Delta\rho$ [see Eq.~(\ref{drho})] in terms of $\overline m_t(m_t)$,
where $\overline m_t(\mu)$ is the top-quark $\overline{\rm MS}$ mass at
renormalization scale $\mu$, and then relates $\overline m_t(m_t)$ to $m_t$
by optimizing the expansion of $m_t/\overline m_t(m_t)$, which is known
through $\O(\alpha_s^2)$ [see Eq.~(\ref{msb})] \cite{gra,lrs},
according to the BLM criterion \cite{blm}.
In Ref.~\cite{alb}, he refines this argument by using the new results of
Ref.~\cite{avd} and an expansion of $\mu_t/\overline m_t(m_t)$, where
$\mu_t=\overline m_t(\mu_t)$, and obtains $\xi=0.323$.
He argues that
an important advantage of this approach is that the coefficients of
$(\alpha_s/\pi)^n$ ($n=1,2$) in the final expressions are either very small
or of $\O(1)$.
In particular, when $\Delta\rho$ is expressed in terms of
$\left[\overline m_t(m_t)\right]^2$, the QCD correction is only
(2--3)${}\cdot10^{-3}$ in the next-to-leading-order approximation.
As a consequence, when $\Delta\rho$ is rewritten in terms of $m_t^2$,
the QCD correction to $\Delta\rho$ is almost entirely contained in
$\left[\overline m_t(m_t)/m_t\right]^2$, i.e., it is a pure QCD effect.
Finally, we observe that the $\O(\alpha_s^2G_Fm_t^2)$ term indeed has the
very sign predicted by the study \cite{kns,fan,nut} of the $t\bar t$
threshold effects and accounts also for the bulk of their size.
In fact, this term may be absorbed into the $\O(\alpha_sG_Fm_t^2)$ term by
choosing $\xi=0.348$ for $N_F=6$ \cite{avd}.
Arguing that $N_F=5$ is more appropriate for $\mu<m_t$, this value comes down
to $\xi=0.324$ \cite{joc}, which is not far outside the range
$0.133\le\xi\le0.287$ predicted from the $t\bar t$ threshold analysis.
The residual difference may be understood by observing that the ladder
diagrams of $\O(\alpha_s^nG_Fm_t^2)$, with $n\ge3$, and the effect of the
finite top-quark decay width are not included in the fixed-order calculation
of Ref.~\cite{avd}.

The claim \cite{ynd} that the $t\bar t$ threshold effects are greatly
overestimated in Refs.~\cite{kns,fan} is based on a simplified analysis,
which demonstrably \cite{nut} suffers from a number of severe analytical and
numerical errors.
Speculations \cite{ghv} that the dispersive computation of $t\bar t$ threshold
effects is unstable are quite obviously unfounded, since they arise from
uncorrelated and unjustifiably extreme variations of the continuum and
threshold contributions.
In particular, the authors of Ref.~\cite{ghv} ascribe the unavoidable scale
dependence of the $\O(\alpha_sG_Fm_t^2)$ continuum result to the uncertainty
in the much smaller threshold contribution, which artificially amplifies
this uncertainty.
In fact, the sum of both contributions, which is the physically relevant
quantity, is considerably less $\mu$ dependent than the continuum
contribution alone \cite{nut}.

\subsection{Specific Corrections: $\Gamma\left(Z\to b\bar b\right)$ and
$\Gamma(Z\to{\rm hadrons})$}

In the so-called improved Born approximation \cite{hol,hal},
$\Gamma\left(Z\to f\bar f\right)$ is given by the formula
\begin{equation}
\label{iba}
\Gamma\left(Z\to f\bar f\right)={N_fG_FM_Z^3\over24\pi\sqrt2}\rho
\left[(1-4|Q_f|s_w^2\kappa)^2R^V+R^A\right],
\end{equation}
where $N_f=1$~(3) for lepton (quark) flavours,
$\rho=(1-\Delta\rho)^{-1}$ and $\kappa=1+\Delta\kappa$, and $R^V$ and $R^A$
are the vector and axial-vector current correlators of $f$.
Keeping the full $m_f$ dependence and including $\O(\alpha)$ electromagnetic
corrections,
the latter read \cite{sch}
\begin{eqnarray}
R^V&\n=\n&\left(1+{1\over2r}\right)\sqrt{1-{1\over r}}
+{3\alpha\over4\pi}Q_f^2\left[1+{3\over r}
+\O\left({\ln r\over r^2}\right)\right],\nonumber\\
R^A&\n=\n&\left(1-{1\over r}\right)^{3/2}
+{3\alpha\over4\pi}Q_f^2\left\{1+{3\over r}\left[\ln(4r)-{1\over2}\right]
+\O\left({\ln r\over r^2}\right)\right\},
\end{eqnarray}
where $r=(M_Z^2/4m_f^2)$.
The QCD corrections to $R^V$ and $R^A$, which are present if $f$ represents a
quark, will be discussed in Sects.~2.2.3 and 2.2.4.
The observable $\Gamma\left(Z\to b\bar b\right)$ deserves special attention,
since it receives specific $m_t$ power corrections.
These may be accommodated in Eq.~(\ref{iba}) by replacing $\rho$ and $\kappa$
with $\rho_b=\rho(1+\tau)^2$ and $\kappa_b=\kappa(1+\tau)^{-1}$, respectively,
where $\tau$ is an additional electroweak parameter \cite{akh}.
Similarly to $\Delta\rho$, $\tau$ receives contributions in the orders
$G_Fm_t^2$, $G_F^2m_t^4$, $\alpha_sG_Fm_t^2$, etc.
Equation~(\ref{iba}) may be rendered exact by including in $\rho$ and $\kappa$
the subleading flavour-dependent weak corrections \cite{akh}.

\vglue 0.3cm
\leftline{\it 2.2.1 Two-Loop $\O(G_F^2m_t^4)$ Corrections for
$M_H\ne0$}
\vglue 1pt

In the oblique corrections considered so far, the $m_t$ dependence
might be masked by all kinds of physics beyond the standard model.
Contrariwise, in the case of $Z\to b\bar b$,
the virtual top quark is tagged directly by the external bottom flavour.
At one loop, there is a strong cancellation between the flavour-independent
oblique corrections, $\Delta\rho$ and $\Delta\kappa$, and the specific
$Z\to b\bar b$ vertex correction, $\tau$ \cite{akh}.

Recently, the leading two-loop corrections to $\tau$,
of $\O(G_F^2m_t^4)$ \cite{bar} and $\O(\alpha_sG_Fm_t^2)$ \cite{fle},
have become available.
The master formula reads
\begin{equation}
\label{tau}
\tau=-2x_t\left[1
+x_t\tau^{(2)}\left({M_H\over m_t}\right)
-2\zeta(2){\alpha_s(m_t)\over\pi}\right],
\end{equation}
where $x_t$ is defined above Eq.~(\ref{drho}).
$\tau^{(2)}(r)$ \cite{bar} rapidly varies with $r$,
$\tau^{(2)}(r)\ge\tau^{(2)}(1.55)=1.23$, and its
asymptotic behaviour is given by
\begin{equation}
\tau^{(2)}(r)=
\cases{
\displaystyle
-2\zeta(2)+9-4\pi r+\O(r^2\ln r),
&if $r\ll1;$\cr
\displaystyle
{5\over2}\ln^2r-{47\over12}\ln r+\zeta(2)+{311\over144}
+O\left({\ln^2r\over r^2}\right),
&if $r\gg1.$\cr}
\end{equation}
The value at $r=0$ has been confirmed by a third group \cite{den}.

\vglue 0.3cm
\leftline{\it 2.2.2 Two-Loop $\O(\alpha_sG_Fm_t^2)$ Corrections}
\vglue 1pt

In Eq.~(\ref{tau}), we have also included the $\O(\alpha_sG_Fm_t^2)$ term
\cite{fle}.
Notice that we have written it in a form appropriate for use in
Eq.~(\ref{iba}) together with the QCD-corrected expressions for $R^V$ and
$R^A$, which have the common beginning  $(1+\alpha_s/\pi)$ (see Sect.~2.2.3).
We observe that the $\O(G_F^2m_t^4)$ and $\O(\alpha_sG_Fm_t^2)$ terms of
Eq.~(\ref{tau}) cancel partially.

\vglue 0.3cm
\leftline{\it 2.2.3 Three-Loop $\O(\alpha_s^3)$ Corrections}
\vglue 1pt

Most of the results discussed in this section are valid also for the
$Z\to q\bar q$ decays with $q\ne b$.
Here, we put $m_q=0$, except for $q=t$.
Finite-$m_b$ effects will be considered in Sect.~2.2.4.
By the optical theorem, the QCD corrections to
$\Gamma\left(Z\to q\bar q\right)$
may be viewed as the imaginary parts of the $Z$-boson self-energy diagrams
that contain a $q$-quark loop decorated with virtual gluons and possibly
other quark loops.
Diagrams where the two $Z$-boson lines are linked to the same quark loop
are usually called non-singlet, while the residual diagrams are called
singlet, which includes the double-triangle diagrams.
By $\gamma_5$ reflection, the non-singlet contribution, $R_{NS}$, to $R^A$
coincides with the one to $R^V$.
Up to $\O(\alpha_s^3)$ in the $\overline{\rm MS}$ scheme with $N_F=5$ active
flavours, one has
\begin{equation}
\label{rns}
R_{NS}=1+{\alpha_s\over\pi}+\left({\alpha_s\over\pi}\right)^2
\left[R_1+F\left({M_Z\over4m_t^2}\right)\right]
+R_2\left({\alpha_s\over\pi}\right)^3,
\end{equation}
where $R_1=1.98571-0.11530\,N_F=1.40923$ \cite{ckt} and
$R_2=-6.63694-1.20013\,N_F-0.00518\,N_F^2=-12.76706$ \cite{sur}.
$F$ collects the decoupling-top-quark effects in $\O(\alpha_s^2)$ and has the
expansion \cite{che}
\begin{equation}
F(r)=r\left[-{8\over135}\ln(4r)+{176\over675}\right]+\O(r^2).
\end{equation}
$F$ has also been obtained in numerical form recently \cite{sop}.
We note that an analytic expression for $F$ had been known previously from
the study of the two-loop QED vertex correction due to virtual heavy
fermions \cite{ver} (see also Ref.~\cite{hoa}).
Recently, the $\O(\alpha_s^4)$ term of Eq.~(\ref{rns}) has been estimated
using the principle of minimal sensitivity and the effective-charges
approach \cite{kat}.
The $\O(\alpha^2)$ and $\O(\alpha\alpha_s)$ corrections to
$\Gamma\left(Z\to b\bar b\right)$ from photonic source are well under
control \cite{alk}.

Due to Furry's theorem \cite{fur}, singlet diagrams with $q\bar qZ$ vector
couplings occur just in $\O(\alpha_s^3)$.
They contain two quark loops at the same level of hierarchy, which,
in general, involve different flavours.
Thus, they cannot be assigned unambiguously to a specific $q\bar q$ channel.
In practice, this does not create a problem, since their combined contribution
to $\Gamma(Z\to{\rm hadrons})$ is very small anyway \cite{sur},
\begin{equation}
\label{rsv}
\delta\Gamma_Z={G_FM_Z^3\over8\pi\sqrt2}\left(\sum_{q=u,d,s,c,b}v_q\right)^2
(-0.41318)\left({\alpha_s\over\pi}\right)^3,
\end{equation}
where $v_q=2I_q-4Q_qs_w^2$.

Axial-type singlet diagrams contribute already in $\O(\alpha_s^2)$.
The sum over triangle subgraphs involving mass-degenerate (e.g., massless)
up- and down-type quarks vanishes.
Thus, after summation, only the double-triangle diagrams involving top and
bottom quarks contribute to $\Gamma\left(Z\to b\bar b\right)$ and
$\Gamma(Z\to{\rm hadrons})$.
The present knowledge of the singlet part, $R_S^A$, of $R^A$ is summarized by
\begin{equation}
\label{rsa}
R_S^A=\left({\alpha_s\over\pi}\right)^2{1\over3}
I\left({M_Z^2\over4m_t^2}\right)+\left({\alpha_s\over\pi}\right)^3
\left({23\over12}\ln^2{m_t^2\over M_Z^2}-{67\over18}\ln{m_t^2\over M_Z^2}
-15.98773\right).
\end{equation}
An analytic expression for the $I$ function may be found in Ref.~\cite{kk};
its high-$m_t$ expansion reads \cite{kk}
\begin{equation}
\label{iexp}
I(r)=3\ln(4r)-{37\over4}+{28\over27}r+\O(r^2).
\end{equation}
The second and third terms on the right-hand side of Eq.~(\ref{iexp}) have
been confirmed recently in Refs.~\cite{kgc,ckw}, respectively.
The $\O(\alpha_s^3)$ logarithmic terms of Eq.~(\ref{rsa}) follow from
Eq.~(\ref{iexp}) by means of renormalization-group techniques \cite{kgc},
while the constant term requires a separate computation \cite{lar}.
Note that, in contrast to the formulations of Ref.~\cite{lar}, Eq.~(\ref{rsa})
is written in terms of the top-quark pole mass.

In $\O(\alpha_s^3)$, $\Gamma(Z\to{\rm hadrons})$ receives a contribution also
from the $Z\to ggg$ decay, which proceeds through quark triangle and box
diagrams.
The decay mode $Z\to gg$ is prohibited by the Landau-Yang theorem \cite{lan}.
The vector and axial-vector parts of $\Gamma(Z\to ggg)$ may be found in
Refs.~\cite{bai,lee}, respectively.
Notice that these contributions are already contained in
Eqs.~(\ref{rsv},\ref{rsa}), respectively.
Less than three out of one million hadronic $Z$-boson decays are due to the
$Z\to ggg$ mechanism \cite{lee}.

\vglue 0.3cm
\leftline{\it 2.2.4 Finite-$m_b$ Effects}
\vglue 1pt

In $\O(\alpha_s)$, the full $m_b$ dependence of $R^V$ and $R^A$ is
known \cite{cgn,sch}, while, in higher orders, only the first terms of their
$m_b^2/M_Z^2$ expansions have been calculated \cite{sgg,chk,ckk}.
In the $\overline{\rm MS}$ scheme with $N_F=5$, one has
\begin{eqnarray}
\label{drv}
\delta R^V&\n=\n&12{\alpha_s\over\pi}\,{\overline m_b^2\over M_Z^2}\left[1+
{629\over72}\,{\alpha_s\over\pi}+45.14610\left({\alpha_s\over\pi}\right)^2
\right],\\
\label{dra}
\delta R^A&\n=\n&-6{\overline m_b^2\over M_Z^2}\left[1+{11\over3}\,
{\alpha_s\over\pi}
+\left({\alpha_s\over\pi}\right)^2\left(11.28560-\ln{m_t^2\over M_Z^2}\right)
\right]\nonumber\\
&&-{10\over27}\left({\alpha_s\over\pi}\right)^2{\overline m_b^2\over m_t^2}
\left({8\over3}+{1\over2}\ln{m_t^2\over M_Z^2}\right),
\end{eqnarray}
where $\alpha_s$ and the bottom-quark $\overline{\rm MS}$ mass,
$\overline m_b$, are to be evaluated at $\mu=M_Z$.
Due to the use of $\overline m_b(M_Z)$ instead of $m_b$,
Eqs.~(\ref{drv},\ref{dra}) are devoid of terms involving $\ln(M_Z^2/m_b^2)$.
The second and third terms of Eq.~(\ref{drv}) come from
Refs.~\cite{sgg,chk},\footnote{There is an obvious typographical error on the
right-hand side of the last equation of Eq.~(23) in Ref.~\cite{chk}:
92 should be replaced by 96.
Furthermore, the coefficient of $\zeta(3)$ in the expression for $b_3$
should read 992/72 instead of 1008/72 \cite{err}.}
respectively, and the $\O(\alpha_s^2)$ terms in the first and second lines of
Eq.~(\ref{dra}) are from Refs.~\cite{ckk,ckw}, respectively.
The $\O(\alpha_s^3\overline m_b^2/M_Z^2)$ term of $R^A$ has not yet appeared in
the literature.
The $\O(\alpha_sm_b^2/M_Z^2)$ corrections should be detectable.
For $m_t=174$~GeV, the finite-$m_b$ terms beyond $\O(\alpha_s)$ in
Eqs.~(\ref{drv},\ref{dra}) each amount to approximately $5\cdot10^{-3}\%$ of
$\Gamma\left(Z\to b\bar b\right)$ but have opposite signs.
As might be expected, the $\O(\alpha_s^2\overline m_b^2/m_t^2)$ term is
insignificant numerically, reducing $\Gamma\left(Z\to b\bar b\right)$
by $3\cdot10^{-5}\%$.

Note that Eqs.~(\ref{drv},\ref{dra}) do not include the contribution due to
the $Z\to q\bar qb\bar b$ decays,\footnote{It is justified to distinguish
between the $Z\to q\bar qb\bar b$ and $Z\to b\bar bq\bar q$ decays, since the
interference of the respective sets of diagrams vanishes upon summation over
isodoublets of $q$ quarks \cite{kk}.} where a bottom-quark pair is radiated
off via a virtual gluon from a primary $q$-quark line, with $q=u,d,s,c$.
These contributions are usually ascribed to the $\O(\alpha_s^2)$ corrections
to $\Gamma\left(Z\to q\bar q\right)$.
Under favourable experimental conditions, these events can be tagged by their
$b\bar b$ invariant mass, which is typically low.
A closed expression for $\Gamma\left(Z\to q\bar qb\bar b\,\right)$ in the
approximation $m_q=0$ may be found in Ref.~\cite{hoa}.
By including also the $q\bar qZ$ vertex correction induced by a virtual bottom
quark \cite{ver}, one obtains the corresponding $\O(\alpha_s^2)$ contribution
to $R_{NS}$ in the $q\bar q$ channel.
It has the expansion \cite{hoa} (see also Ref.~\cite{sop})
\begin{equation}
\label{qqbb}
\delta R_{NS}=\left({\alpha_s\over\pi}\right)^2\left\{{2\over3}\zeta(3)
-{11\over12}+x^2\left[-\ln x-4\zeta(3)+{13\over3}\right]+\O(x^3)\right\},
\end{equation}
where $x=m_b^2/M_Z^2$.
The first term on the right-hand side of Eq.~(\ref{qqbb}) coincides with the
coefficient of the $N_F$-dependent term of $R_1$ in Eq.~(\ref{rns}).
Notice that the logarithmic term in Eq.~(\ref{qqbb}) cannot be eliminated
by introducing $\overline m_b(M_Z)$.
Due to the absence of a term linear in $x$, the finite-$m_b$ corrections
in Eq.~(\ref{qqbb}) will be difficult to probe experimentally.

For the reader's convenience, we recall the procedure of evaluating
$\overline m_b(\mu)$.
For later use in Sect.~3.2.1, we consider also charm quarks.
The quark pole masses, $m_q$, may be estimated from QCD sum rules \cite{sum}.
Recent analyses \cite{pol} have yielded $m_c=(1.46\pm0.07)$~GeV and
$m_b=(4.72\pm0.05)$~GeV.
To obtain $\overline m_q(\mu)$, one proceeds in two steps.
Firstly, one evaluates $\overline m_q(m_q)$ from
\begin{equation}
\label{msb}
{m_q\over\overline m_q(m_q)}=1+C_F{\alpha_s(m_q)\over\pi}
+K_q\left[{\alpha_s(m_q)\over\pi}\right]^2,
\end{equation}
with \cite{gra,lrs}
\begin{equation}
K_q\approx16.11006-1.04137\sum_{i=1}^{N_F-1}\left(1-{4\over3}\,
{m_i\over m_q}\right),
\end{equation}
where the sum extents over all quark flavours with $m_i<m_q$.
Assuming $m_s=0.266$~GeV \cite{dom} and neglecting $m_u$ and $m_d$, one
finds $K_c=13.2$ and $K_b=12.5$.
Secondly, one determines $\overline m_q(\mu)$ via the scaling law
\begin{equation}
\overline m_q(\mu)=\overline m_q(m_q)
{c_q(\alpha_s(\mu)/\pi)\over c_q(\alpha_s(m_q)/\pi)},
\end{equation}
where \cite{gra,gor}
\begin{eqnarray}
c_c(x)&\n=\n&\left({25\over6}x\right)^{12/25}(1+1.01413x+1.38921x^2),\nonumber\\
c_b(x)&\n=\n&\left({23\over6}x\right)^{12/23}(1+1.17549x+1.50071x^2).
\end{eqnarray}

\section{Higgs Sector: Corrections to $\Gamma\left(H\to f\bar f\,\right)$ and
$\Gamma(H\to{\rm hadrons})$}

Quantum corrections to Higgs-boson phenomenology have received
much attention in the literature; for a review, see Ref.~\cite{bak}.
The experimental relevance of radiative corrections to the $f\bar f$
branching fractions of the Higgs boson has been emphasized recently
in the context of a study dedicated to LEP~2 \cite{gkw}.
Techniques for the measurement of these branching fractions at a
$\sqrt s=500$~GeV $e^+e^-$ linear collider have been elaborated in
Ref.~\cite{hil}.

In the Born approximation, the $f\bar f$ partial widths of the Higgs boson
are given by \cite{res}
\begin{equation}
\label{born}
\Gamma_0\left(H\to f\bar f\,\right)={N_fG_FM_Hm_f^2\over4\pi\sqrt2}
\left(1-{4m_f^2\over M_H^2}\right)^{3/2}.
\end{equation}
The full one-loop electroweak corrections to Eq.~(\ref{born}) are now well
established \cite{jfl,hff}.
They consist of an electromagnetic and a weak part, which are separately
finite and gauge independent.
They may be included in Eq.~(\ref{born}) as an overall factor,
$\left[1+(\alpha/\pi)Q_f^2\Delta_{em}\right](1+\Delta_{weak})$.
For $M_H\gg2m_f$, $\Delta_{em}$ develops a large logarithm,
\begin{equation}
\label{delem}
\Delta_{em}=-{3\over2}\ln{M_H^2\over m_f^2}+{9\over4}
+\O\left({m_f^2\over M_H^2}\ln{M_H^2\over m_f^2}\right).
\end{equation}
For $M_H\ll2M_W$, the weak part is well approximated by \cite{hff}
\begin{equation}
\label{weak}
\Delta_{weak}={G_F\over8\pi^2\sqrt2}\left\{C_fm_t^2
+M_W^2\left({3\over s_w^2}\ln c_w^2-5\right)
+M_Z^2\left[{1\over2}-3\left(1-4s_w^2|Q_f|\right)^2\right]\right\},
\end{equation}
where $C_b=1$ and $C_f=7$ for all other flavours, except for top.
The $t\bar t$ mode will not be probed experimentally anytime soon
and we shall not be concerned with it in the remainder of this article.
Equation~(\ref{weak}) has been obtained by putting $M_H=m_f=0$
($f\ne t$) in the expression for the full one-loop weak correction.
It provides a very good approximation for $f=\tau$ up to
$M_H\approx135$~GeV and for $f=b$ up to $M_H\approx70$~GeV,
the relative deviation from the full weak correction being less than 15\%
in each case.

\subsection{Two-Loop $\O(\alpha_sG_Fm_t^2)$ Corrections to
$\Gamma\left(H\to f\bar f\,\right)$}

{}From Eq.~(\ref{weak}) it is evident that the dominant effect is due to
virtual
top quarks.
In the case $f\ne b$, the $m_t$ dependence is carried solely by the
renormalizations of the wave function and the vacuum expectation value
of the Higgs field and is thus flavour independent.
These corrections are of the same nature as those considered in
Ref.~\cite{cha}.
For $f=b$, there are additional $m_t$-dependent contributions from the
$b\bar bH$ vertex correction and the bottom-quark wave-function
renormalization.
Incidentally, they cancel almost completely the universal $m_t$ dependence.
It is amusing to observe that a similar situation has been encountered in the
context of the $Z\to f\bar f$ decays \cite{akh}.
The QCD corrections to the universal and non-universal $\O(G_Fm_t^2)$ terms
will be presented in Sects.~3.1.1 and 3.1.2, respectively.

\vglue 0.3cm
\leftline{\it 3.1.1 Universal Corrections}
\vglue 1pt

The universal $\O(G_Fm_t^2)$ term of $\Delta_{weak}$ resides inside the
combination
\begin{equation}
\label{delta}
\Delta_u=-{\Pi_{WW}(0)\over M_W^2}-\re \Pi_{HH}^\prime\left(M_H^2\right),
\end{equation}
where $\Pi_{WW}$ and $\Pi_{HH}$ are the unrenormalized self-energies
of the $W$ and Higgs bosons, respectively \cite{hff}.
The same is true of its QCD correction.

For $M_H<2m_t$ and $m_b=0$, the one-loop term reads \cite{hff}
\begin{equation}
\label{delzero}
\Delta_u^0=4N_cx_t\left[\left(1+{1\over2r}\right)
\sqrt{{1\over r}-1}\arcsin\sqrt r-{1\over4}-{1\over2r}\right],
\end{equation}
where $r=(M_H^2/4m_t^2)$ and $x_t$ is defined above Eq.~(\ref{drho}).
In the same approximation, the two-loop term may be written as \cite{hll,two}
\begin{equation}
\label{delone}
\Delta_u^1=N_cC_Fx_t{\alpha_s\over\pi}
\left[6\zeta(3)+2\zeta(2)-{19\over4}-\re H_1^\prime(r)\right],
\end{equation}
where $C_F=\left(N_c^2-1\right)/(2N_c)=4/3$ and $H_1$ has an expression in
terms of dilogarithms and trilogarithms \cite{two}.
Equation~(\ref{delone}) has been confirmed recently \cite{djg}.
In the heavy-quark limit ($r\ll1$), one has \cite{two}
\begin{equation}
H_1^\prime(r)=6\zeta(3)+3\zeta(2)-{13\over4}+{122\over135}r+\O(r^2).
\end{equation}
Combining Eqs.~(\ref{delzero},\ref{delone}) and retaining only the leading
high-$m_t$ terms, one finds the QCD-corrected coefficients $C_f$ for $f\ne b$,
\begin{equation}
\label{kun}
C_f=7-2\left({\pi\over3}+{3\over\pi}\right)\alpha_s
\approx7-4.00425\,\alpha_s.
\end{equation}
This result has been reproduced recently \cite{kwi}.
We recover the notion that, in electroweak physics, the one-loop
$\O\left(G_Fm_t^2\right)$ terms get screened by their QCD corrections.
The QCD correction to the shift in $\Gamma\left(H\to f\bar f\,\right)$
induced by a pair of novel quarks with arbitrary masses may be found in
Ref.~\cite{two}.

\vglue 0.3cm
\leftline{\it 3.1.2 Non-universal Corrections}
\vglue 1pt

The QCD correction to the non-universal one-loop contribution to
$\Gamma\left(H\to b\bar b\right)$ arises in part from genuine two-loop
three-point diagrams, which are more involved technically.
However, the leading high-$m_t$ term may be extracted \cite{spi} by means of a
low-energy theorem \cite{ell},
which relates the amplitudes of two processes that differ by the
insertion of an external Higgs-boson line carrying zero momentum.
In this way, one only needs to compute the irreducible two-loop bottom-quark
self-energy diagrams with one gluon and one longitudinal $W$ boson, which
may be taken massless.
After using the Dirac equation and factoring out one power of $m_b$,
one may put $m_b=0$ in the two-loop integrals,
which may then be solved analytically.
Applying the low-energy theorem and performing on-shell renormalization,
one eventually finds the non-universal leading high-$m_t$ term along with
its QCD correction \cite{spi},
\begin{equation}
\Delta_{nu}=x_t\left(-6+{3\over2}C_F{\alpha_s\over\pi}\right).
\end{equation}
Combining the term contained within the parentheses with Eq.~(\ref{kun}),
one obtains the QCD-corrected coefficient $C_b$,
\begin{equation}
\label{knu}
C_b=1-2\left({\pi\over3}+{2\over\pi}\right)\alpha_s
\approx1-3.36763\,\alpha_s.
\end{equation}
Again, the $\O(G_Fm_t^2)$ term is screened by its QCD correction.

\subsection{QCD Corrections to $\Gamma(H\to{\rm hadrons})$}

Both $H\to q\bar q$ and $H\to gg$ lead to two-jet final states.
These processes are relevant phenomenologically for the search for the
low- and intermediate-mass Higgs boson, with $M_H<2M_W$, at $e^+e^-$ colliders,
while they are very difficult to discern from background reactions at hadron
colliders.
The partial widths of both decay modes receive significant QCD corrections.

\vglue 0.3cm
\leftline{\it 3.2.1 Two-Loop $\O(\alpha_s^2)$ Corrections to
$\Gamma\left(H\to q\bar q\right)$ Including Finite-$m_q$ Effects}
\vglue 1pt

In the on-shell scheme, the one-loop QCD correction \cite{bra} to
$\Gamma\left(H\to q\bar q\right)$ emerges from one-loop QED correction by
substituting $\alpha_sC_F$ for $\alpha Q_f^2$.
{}From Eq.~(\ref{delem}) it is apparent that, for $m_q\ll M_H/2$, large
logarithmic corrections occur.
In general, they are of the form $(\alpha_s/\pi)^n\ln^m(M_H^2/m_q^2)$,
with $n\ge m$.
Owing to the renormalization-group equation, these logarithms may be
absorbed completely into the running $\overline{\rm MS}$ quark mass,
$\overline m_q(\mu)$, evaluated at $\mu=M_H$.
A similar mechanism has been exploited also in Eqs.~(\ref{drv},\ref{dra}).
In this way, these logarithms are resummed to all orders and the perturbation
expansion converges more rapidly.
This observation gives support to the notion that the $q\bar qH$ Yukawa
couplings are controlled by the running quark masses.

For $q\ne t$, the QCD corrections to $\Gamma\left(H\to q\bar q\right)$ are
known up to $\O(\alpha_s^2)$.
In the $\overline{\rm MS}$ scheme, the result is \cite{bak}
\begin{eqnarray}
\label{hqqmsb}
\Gamma\left(H\to q\bar q\right)&\n=\n&{3G_FM_H\overline m_q^2\over4\pi\sqrt2}
\left\{\left(1-4{\overline m_q^2\over M_H^2}\right)^{3/2}
+C_F{\alpha_s\over\pi}\left({17\over4}-30{\overline m_q^2\over M_H^2}\right)
\right.\nonumber\\
&&\qquad{}+\left({\alpha_s\over\pi}\right)^2\left[K_1
+K_2{\overline m_q^2\over M_H^2}+12\sum_{i=u,d,s,c,b}
{\overline m_i^2\over M_H^2}
\right.\nonumber\\
&&\qquad{}+\left.\left.
{1\over3}\left({1\over3}\ln^2{M_H^2\over\overline m_q^2}
-2\ln{M_H^2\over m_t^2}+8-2\zeta(2)\right)
\vphantom{\left(1-4{\overline m_q^2\over M_H^2}\right)^{3/2}}\right]\right\},
\end{eqnarray}
where $K_1=35.93996-1.35865\,N_F$ \cite{gor},
$K_2=-129.72924+6.00093\,N_F$ \cite{lrs}, with $N_F$ being the number
of quark flavours active at $\mu=M_H$, and it is understood that $\alpha_s$,
$\overline m_q$, and $\overline m_i$ are to be evaluated at this scale.
The last term in the second line of Eq.~(\ref{hqqmsb}) has been found in
Ref.~\cite{lrs}, too.
The term in the third line is related to the double-triangle diagram involving
the $q$ and top quarks \cite{cak};
its logarithmic dependence on $\overline m_q$ is compensated in the
Higgs-boson hadronic width by a similar term connected with the $gg$ final
state \cite{cak}.
Equation~(\ref{hqqmsb}) may be translated into the on-shell scheme by using
the relation between $m_q$ and $\overline m_q(M_H)$ (see Sect.~2.2.4)
\cite{kim}.
The difference between these two evaluations is extremely small \cite{kim},
which indicates that the residual uncertainty due to the lack of knowledge of
the $\O(\alpha_s^3)$ term is likely to be inconsequential for practical
purposes.

The electroweak corrections may be implemented in Eq.~(\ref{hqqmsb}) by
multiplication with
$\left[1+(\alpha/\pi)Q_f^2\Delta_{em}\right](1+\Delta_{weak})$,
where $\Delta_{em}$ and $\Delta_{weak}$ are given in
Eqs.~(\ref{delem},\ref{weak}), respectively.
To include also the $\O(\alpha_sG_Fm_t^2)$ corrections, one substitutes in
Eq.~(\ref{weak}) the QCD-corrected $C_f$ terms specified in
Eqs.~(\ref{kun},\ref{knu}).
We note in passing that our result \cite{spi} disagrees with a recent
calculation \cite{kwi} of the $\O(\alpha_sG_Fm_t^2)$ correction to
$\Gamma\left(H\to b\bar b\right)$ in the on-shell scheme.

\vglue 0.3cm
\leftline{\it 3.2.2 Two-Loop $\O(\alpha_s)$ Corrections to $\Gamma(H\to gg)$}
\vglue 1pt

The hadronic width of the Higgs boson receives contributions also from
the $H\to gg$ channel, which is mediated by massive-quark triangle diagrams.
To lowest order, $\Gamma(H\to gg)$ \cite{wil} is well approximated by
\cite{bak}
\begin{equation}
\label{hgg}
\Gamma(H\to gg)={\alpha_s^2G_FM_H^3\over36\pi^3\sqrt2}
\left(1+{7\over60}\,{M_H^2\over m_t^2}\right).
\end{equation}
The next-to-leading-order QCD correction may be accommodated in
Eq.~(\ref{hgg}) by multiplication with \cite{adj}
\begin{equation}
\label{hggg}
\left[1+{\alpha_s(M_H)\over\pi}\left({95\over4}-{7\over6}N_F\right)\right],
\end{equation}
where $N_F$ is the number of quark flavours active at scale $M_H$.
Equation~(\ref{hggg}) includes two-loop corrections in the $gg$ channel and
one-loop contributions due to the $ggg$ and $gq\bar q$ modes, where the
$q\bar q$ pair is created from a virtual gluon.
Assuming $N_F=5$ and $\alpha_s(M_H)=0.11$, Eq.~(\ref{hggg}) is as large as
1.63.
For $M_H<2M_H$, the two-loop QCD correction to $\Gamma(H\to\gamma\gamma)$ is
below 3\% in size \cite{zhe} and the one to $\Gamma(H\to Z\gamma)$ below
0.3\% \cite{msp}.

\section{Conclusions}

In conclusion, all dominant two-loop and even certain three-loop radiative
corrections to $Z$-boson physics are now available.
However, one has to bear in mind that, apart from the lack of knowledge of
the accurate values of $M_H$, $m_t$, and $\alpha_s(M_Z)$, the reliability
of the theoretical predictions is limited by a number of error sources.
The inherent QCD errors on the hadronic contribution to $\Delta\alpha$ and
the $tb$ contribution to $\Delta\rho$ are
$\delta\Delta\alpha=\pm7\cdot10^{-4}$ and
$\delta\Delta\rho=\pm1.5\cdot10^{-4}$, respectively, which amounts to
$\delta\Delta r=\pm8.6\cdot10^{-4}$.
The unknown electroweak corrections are of the order
$(\alpha/\pi s_w^2)^2(m_t^2/M_Z^2)\ln(m_t^2/M_Z^2)\approx6\cdot10^{-4}$,
possibly multiplied by a large prefactor \cite{dfg}.
The scheme dependence of the key electroweak parameters has been estimated
in Refs.~\cite{fan,ber,boc}
by comparing the evaluations in the on-shell
scheme and certain variants of the $\overline{\rm MS}$ scheme;
the maximum variation of $\Delta r$ in the ranges
60~GeV${}\le M_H\le{}$1~TeV and 150~GeV${}\le m_t\le{}$200~GeV
is $8\cdot10^{-5}$ when the coupling-constant renormalization
is converted \cite{fan} and $4\cdot10^{-4}$ when the top-quark mass is
redefined taking into account just the QCD corrections \cite{ber}.
The effect on $\Delta\rho$ of including also the leading electroweak
corrections in the redefinition of the top-quark mass has been
investigated recently in the approximation $M_H,m_t\gg M_Z$ \cite{boc}.
The theoretical predictions for Higgs-boson physics at present and
near-future colliding-beam experiments are probably far more precise than the
expected theoretical errors.

\bigskip
\centerline{\bf ACKNOWLEDGMENTS}
\smallskip\noindent
I would like to thank Jochem Fleischer, Hans K\"uhn, Alberto Sirlin, and
Levan Surguladze for clarifying comments regarding Refs.~\cite{avd},
\cite{hoa,chk,ckk}, \cite{asi,alb}, and \cite{sgg}, respectively.
I am grateful to the KEK Theory Group for the warm hospitality extended to
me during my visit, when this review article was written.
This work was supported by the Japan Society for the Promotion of Science
(JSPS) through Fellowship No.~S94159.

\end{document}